\documentclass[pss]{wiley2sp} 
\usepackage{amsmath,amssymb,bbm}

\begin{document}

\title{Getting back to Na$_x$CoO$_2$:\\
spectral and thermoelectric properties}

\titlerunning{Getting back to Na$_x$CoO$_2$}

\author{%
  L. Boehnke\textsuperscript{\textsf{\bfseries 1}} and
  F. Lechermann\textsuperscript{\Ast,\textsf{\bfseries 1}}}

\authorrunning{L. Boehnke and F. Lechermann}

\mail{Frank.Lechermann@physnet.uni-hamburg.de}

\institute{%
  \textsuperscript{1}\, I. Institut f\"ur Theoretische Physik, 
   Universit\"at Hamburg, Germany}

\received{XXXX, revised XXXX, accepted XXXX} 
\published{XXXX} 


\abstract{%
\abstcol{Sodium cobaltate Na$_x$CoO$_2$ as dopable strongly correlated 
layered material with a triangular sublattice still poses a challenging problem in 
condensed matter. The intriguing interplay between lattice, charge, spin and orbital
degrees of freedom leads to a complex phase diagram bounded by a nominal Mott ($x$=0)
regime and a band-insulating ($x$=1) phase. By means of the charge self-consistent 
density functional theory (DFT) plus dynamical mean-field theory (DMFT) scheme, 
built on a pseudopotential framework combined with a continuous-time
quantum Monte-Carlo solver, we here study the one-particle spectral function 
$A({\bf k},\omega)$ as well as the thermopower $S(T)$.}
{The computations may account for the suppression of the $e_g'$ pockets in 
$A({\bf k},\omega)$ at lower doping in line with photoemission experiments. 
Enhancement of the thermopower is verified within the present elaborate multi-orbital 
method to treat correlated materials. In addition, the two-particle dynamic 
spin susceptibility $\chi_s(\omega,{\bf q})$ is investigated based on a simplified 
tight-binding approach, yet by including vertex contributions in the DMFT linear 
response. Besides the identification of paramagnon branches at higher doping,
a prominent high-energy antiferromagnetic mode close to $x$=0.67 is therewith 
identified in $\chi_s(\omega,{\bf q})$, which can be linked to extended hopping
terms on the CoO$_2$ sublattice.}}

\maketitle   

\section{Introduction}
The quasi two-dimensional sodium cobaltate system Na$_x$CoO$_2$ marks one milestone 
in the investigation of realistic strongly correlated electron systems. 
It consists of stacked triangular 
CoO$_2$ layers, glued together by Na ions inbetween. Depending on the doping $x$,
the nominal oxidation state of the cobalt ion lies between Co$^{4+}$($3d^5$) and 
Co$^{3+}$($3d^6$). While the $x$=1 compound is a band insulator with a filled
low-spin Co($t_{2g}$) subshell, for $x$=0 a single hole resides therein. Stimulated 
by findings of large thermoelectric response at higher doping $x$~\cite{ter97,mot01} 
and superconductivity for $x$$\sim$0.3 upon intercalation with water~\cite{tak03}, the 
phase diagram of Na$_x$CoO$_2$ attracted enormous interest, both experimentally as 
well as theoretically, in the pre-pnictide era of the early 2000 century. The 
relevance of strong correlation effects due to the partially filled 
Co($3d$) shell for $x$$<$1 has been motivated by several experiments, e.g., 
from optics~\cite{wan04}, photoemission~\cite{val02,has04,yan07,gec07} and 
transport~\cite{foo04} measurements.

Although much progress has been made in the understanding of layered cobaltates,
after more than ten years of extensive research many problems are still open. We here 
want to address selected matters of debate, namely the nature of the low-energy 
electronic one-particle spectral function, the pecularities of the dynamic spin 
response as well as the temperature- and doping-dependent behavior of the Seebeck 
coefficient.

\section{Theoretical approach}
Effective single-particle methods based on the local density approximation
(LDA) to density functional theory (DFT) are known to be insufficient to cover the
rich physics of strongly correlated materials. Tailored model-Hamiltonian approaches 
to be treated within a powerful many-body technique such as dynamical mean-field 
theory (DMFT) are thus useful to reveal important insight in the dominant processes
at high and low energy. Nowadays the DFT+DMFT methodology (see 
e.g.~\cite{kotliar_review} for a review) opens the possibility to tackle electronic 
correlations with the benefit of the fully interlaced LDA materials chemistry.

In this work a charge self-consistent DFT+DMFT scheme~\cite{gri12} built up on an 
efficient combination of a mixed-basis pseudopotential framework~\cite{mbpp_code} 
with a hybridization-expansion continuous-time quantum Monte-Carlo 
solver~\cite{rub05,wer06,triqs_code,boe11} is utilized to retrieve spectral functions
and the thermopower. Thereby the correlated subspace consists of the 
projected~\cite{ama08,ani05} $t_{2g}$ orbitals, i.e. a three-orbital many-body 
treatment is performed within the single-site DMFT part. The generic multi-orbital 
Coulomb interactions include density-density as well as spin-flip and 
pair-hopping terms, parametrized~\cite{cas78,fre97} by a Hubbard $U$=5 eV and a 
Hund's exchange $J$=0.7 eV. Since the physics of sodium cobaltate is 
intrinsically doping dependent, we constructed Na pseudopotentials with fractional 
nuclear charge in order to cope therewith in DFT+DMFT. A simplistic structural 
approach was undertaken, utilizing a primitive hexagonal cell allowing for only one 
formula unit of Na$_x$CoO$_2$, with the fractional-charge Na in the so-called Na2 
position. Thus note that therefore the bilayer splitting does not occur in the 
electronic structure. Our calculations are straightforwardly extendable to more 
complex unit cells and geometries, however the present approach suits already
the purpose of allowing for some general qualitative statements.

The resulting 3$\times 3$ DFT+DMFT Green's function in Bloch space with one
correlated Co ion in the primitive cell hence reads here~\cite{gri12}
\begin{eqnarray}
\label{Eqn:GBLDEFINITION}
{\bf G}^{\mathrm{bl}} ({\bf k}, i \omega_n)&=&
\Bigl[ (i \omega_n + \mu) {\bf 1} - \varepsilon_{\bf k}^{\mathrm{KS}}-\nonumber\\
&&-\, {\bf P}^{\dagger}({\bf k}) \cdot{\bf \Delta\Sigma}^{\rm imp}
(i\omega_n)\cdot{\bf P}({\bf k})\Bigr]^{-1}\;,
\end{eqnarray}
where $\varepsilon_{\bf k}^{\mathrm{KS}}$ denotes the Kohn-Sham (KS) dispersion part, 
$\mu$ is the chemical potential and ${\bf P}({\bf k})$ the $t_{2g}$ projection 
matrix mediating between the Co correlated subspace and the crystal Bloch space. The 
impurity self-energy term ${\bf \Delta\Sigma}^{\rm imp}$ includes the DMFT self-energy
modified by the double-counting correction. For the latter the fully-localized 
form~\cite{ani93} has been utilized. To extract the one-particle spectral function 
$A(\vec{k},\omega)$=$-\pi^{-1}\,{\rm Im}\,G^{\mathrm{bl}}({\bf k},\omega)$ 
as well as the thermopower, an analytical continuation of the
impurity self-energy term ${\bf \Delta\Sigma}^{\rm imp}$ in Matsubara space 
$\omega_n$ was performed via  Pad{\'e} approximation. Note that via the
upfolding procedure within eq.~(\ref{Eqn:GBLDEFINITION}), the resulting 
real-frequency self-energy term in Bloch space carries $k$ dependence, i.e. 
${\bf \Delta\Sigma}^{\rm bl}$=${\bf \Delta\Sigma}^{\rm bl}({\bf k},\omega)$.

For the investigation of the thermoelectric response, the Seebeck coefficient is 
calculated within the Kubo formalism from
\begin{equation}
S=-\frac{k_B}{|e|}\frac{A_1}{A_0}\;,
\end{equation}
where the correlation functions $A_n$ are given by
\begin{align}
A_n=\sum_{\vec{k}}\int d\omega&\,\beta^n(\omega-\mu)^n
\left(-\frac{\partial f_\mu}{\partial\omega}\right)\notag\times\\
&\times\operatorname{Tr}\left[\vec{v}(\vec{k})A(\vec{k},\omega)\vec{v}
(\vec{k})A(\vec{k},\omega)\right]\;. \label{eqn:An}
\end{align}
Here $\beta$ is the inverse temperature, $\vec{v}(\vec{k})$ denotes the Fermi 
velocity calculated from the charge 
self-consistent KS part and $f_{\mu}$ marks the Fermi-Dirac distribution. 
Due to subtle refinements in the low-energy regime close to the Fermi level within
the charge self-consistent DFT+DMFT scheme, computing $A_n$ through 
eq.~\eqref{eqn:An} for the three-orbital system at hand is quite challenging. It
requires great care both in the handling of the frequency dependence of the spectral 
function through the analytical continuation of the local self-energy term as well as 
the evaluation of the $k$ sum. The difficulty with the latter is the sharp structure 
of the summand especially for low temperatures. Even a tetrahedron summation is 
problematic due to the double appearance of the spectral function in the 
sum~\cite{pal01}. We overcome this problem by using an adaptive numerical integration 
method separately for each $A_n$, where $\Delta\Sigma^{\rm bl}(\vec{k},\omega)$, 
$\vec{v}(\vec{k})$ and $\varepsilon_{\vec{k}}^{\rm KS}$ are linearly interpolated in 
reciprocal space. Since all these quantities are relatively smooth in $k$, the 
resulting $A_n$ show only weak dependence on the underlying mesh for these 
interpolations.

Finally, the expensive two-particle-based dynamic spin response with 
relevant local vertex corrections was studied. Thereby a simplified single-band 
tight-binding parametrization~\cite{ros03} of the realistic dispersion, including 
hopping integrals up to third-nearest neighbor, entered the DMFT self-consistency 
cycle. For more details on the utilized DMFT+vertex technique see~\cite{boe11,boe12}.
\begin{figure*}[t]%
\begin{center}
\includegraphics*[width=0.45\textwidth]{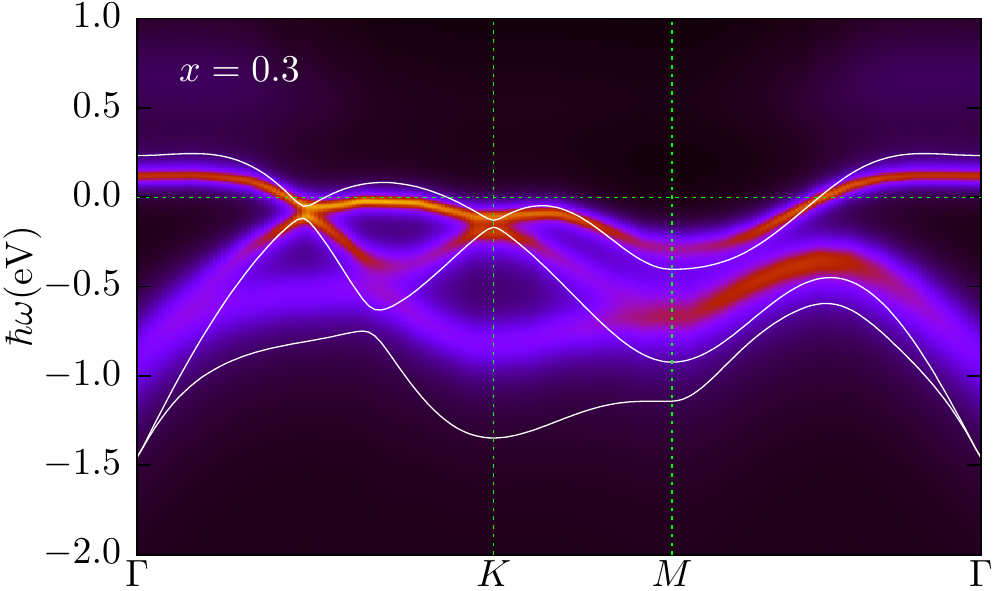}
\hspace*{0.2cm}\vline\hspace*{0.2cm}
\includegraphics*[width=0.45\textwidth]{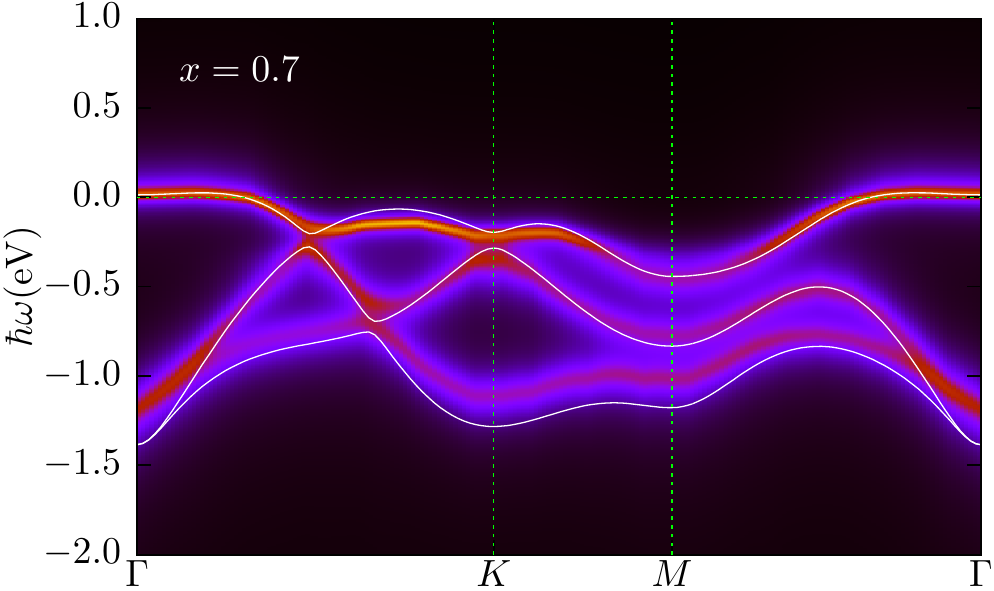}
\end{center}
\caption{DFT+DMFT spectral function $A({\bf k},\omega)$ for $x$=0.3 (left) 
and $x$=0.7 (right) at temperature $T$=290K. For comparison the LDA band structure
is drawn in white. The apical oxygen position was chosen such that an LDA 
single-sheet FS barely exists for $x$=0.7}
\label{ospec}
\end{figure*}
\section{One-particle spectral function}

The low-energy electronic states of Na$_x$CoO$_2$ close to the Fermi level 
$\varepsilon_{\rm F}$ have been subject to many discussions. LDA calculations
for single-formula-unit cells reveal a threefold bandstructure of about 1.5 eV
total bandwidth, dominantly originating from the Co $3d(t_{2g})$ states~\cite{sin00}. 
The resulting LDA Fermi surface (FS) consists of an $a_{1g}$-like hole sheet with 
additional $e_{g}'$-like hole pockets near the $K$ point of the hexagonal 
1. Brillouin zone (BZ). For larger doping these hole pockets become more and more 
filled and their existence for $x$$\gtrsim$0.6 subtly depends on the very structural 
details~\cite{joh04}. Not only displays the measured spectral function 
$A({\bf k},\omega)$ from angle-resolved photoemission (ARPES) experiments a much 
narrower dispersion very close to $\varepsilon_{\rm F}$, but also the FS at lower 
doping lacks the hole-pocket sheets for {\sl any} doping 
$x$~\cite{has04,qia06_2,yan07,gec07}. Usually LDA works surprisingly well for the FS
of strongly correlated metals, even if the method does not allow for the proper
renormalization and the appearance of Hubbard sidebands. Hence sodium cobaltate 
seems to belong to rare cases of correlated metals where the LDA FS topology does 
not agree with experiment.

Many attempts have been elaborated in order to either explain the non-existence of 
the hole pockets or to prove the ARPES data wrong. Without going into the very 
details of this rather long story, no definite final decision has been made on 
either line of argument. Concerning proper correlated methodologies, DFT+DMFT without 
charge self-consistency, i.e. in the traditional post-processing manner, may even 
{\sl increase} the strength of the hole pockets (see e.g.~\cite{mar07}). It appeared 
that the size of the $a_{1g}$-$e_{g}'$ crystal-field splitting plays an important 
role when 
turning on correlations~\cite{lecproc,mar07}. From an LDA+Gutzwiller study by 
Wan {\sl et al.}~\cite{wan08} it became clear that charge self-consistency may has a 
relevant influence on the correlated FS and the hole pockets indeed disappeared for 
$x$$>$0.3 in their work.

In order to touch base with these results we computed the one-particle spectral 
function $A({\bf k},\omega)$ within charge self-consistent DFT+DMFT for $x$=0.3
and $x$=0.7. Thereby within our simplified structural treatment the apical oxygen
position was chosen such to allow for a single-sheet $a_{1g}$-like hole FS within
LDA at $x$=0.7. Though the impact of charge order onto the spectral function is
believed to be also important~\cite{pei11}, we here neglect this influence and 
concentrate on the multi-orbital interplay and its impact on the correlated
Fermi surface. Figure~\ref{ospec} shows the obtained three-orbital spectral functions
close to the Fermi level. Note that there is also a lower Hubbard band, but due to 
the strong doping from half filling it is located in the energy range $[-6,-4]$ eV. As
expected, the less-doped $x$=0.3 case shows a stronger total renormalization of the
$t_{2g}$ derived manifold. The most important observation is the clear shift of the 
potential pocket-forming $e_{g}'$-like quasiparticle bands away from 
$\varepsilon_{\rm F}$ compared to the LDA result. This here amounts to a (nearly 
complete) vanishing of the pockets for $x$=0.3, where they are still sizable in LDA. 
Even if there are some ambiguities concerning possible modifications due to structural
details, the main trend that charge self-consistent DFT+DMFT (notably without
invoking long-range order) tends to surpress the 
$e_{g}'$ derived pockets is evident. Additionally the $e_g'$-like states exhibit a
substantial broadening also with (nearly) total filling, a multi-orbital effect
discussed already for filled $t_{2g}$ states in LaNiO$_3$~\cite{den12}. This 
altogether brings the theoretical description in line with the available 
ARPES data. Thus charge self-consistency can be an important ingredient in the 
calculations, accounting for shifts of the level structure in sensitive crystal-field
environments. 

\section{Transport: Seebeck coefficient}
\begin{figure}[b]%
\begin{center}
\includegraphics*[height=6cm]{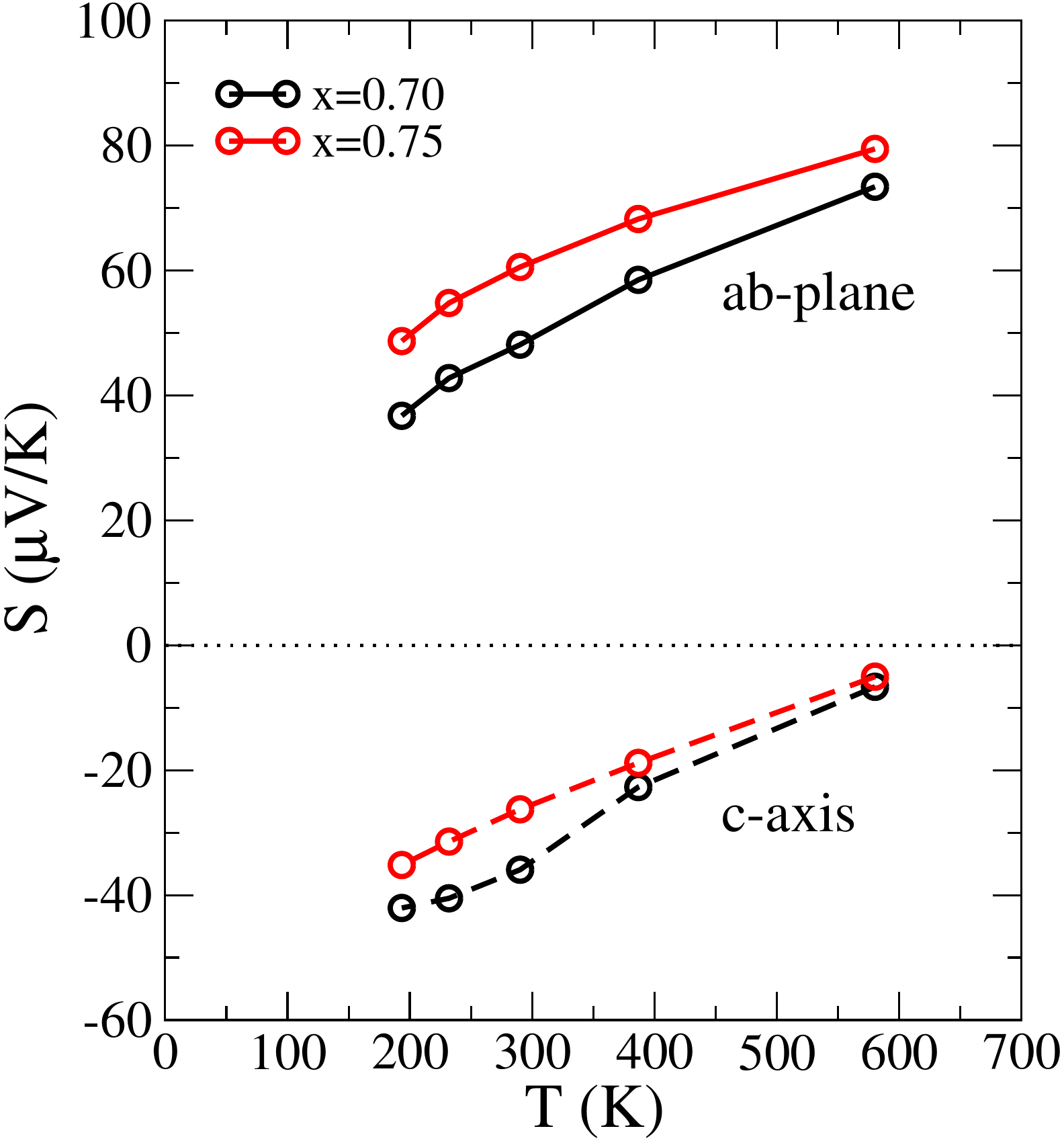}
\end{center}
\caption{Seebeck coefficient $S(T)$ within charge self-consistent DFT+DMFT for 
larger doping. Full lines correspond to the inplane thermopower, dashed lines to 
$S(T)$ along the $c$-axis.}\label{seebeck-pic}
\end{figure}
\begin{figure*}[t]%
\begin{center}
\includegraphics*[height=6cm]{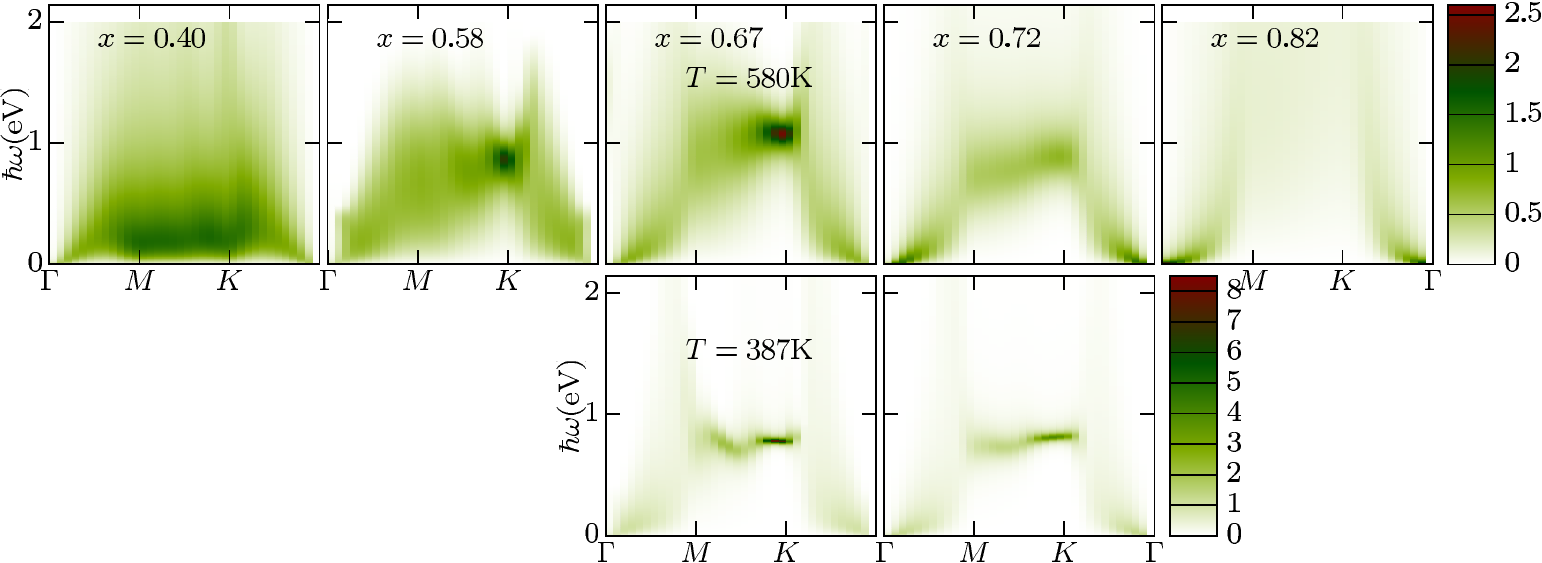}
\end{center}
\caption{Doping-dependent dynamic spin susceptibility $\chi_s(\omega,{\bf q},T)$.
The lower row shows the results for reduced $T$ for the same respective doping as
directly above.}
\label{dynsus}
\end{figure*}
\begin{figure}[t]%
\begin{center}
\includegraphics*[height=4cm]{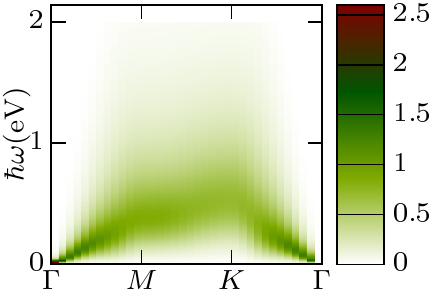}
\end{center}
\caption{Dynamic spin susceptibility $\chi_s(\omega,{\bf q},$$T$=580K$)$ for only
nearest-neighbor hopping $-t$ at $x$=0.67.}
\label{dynsus_tb}
\end{figure}
The increased thermoelectric response at larger doping $x$ marks one of the 
Na$_x$CoO$_2$ key aspects~\cite{ter97,mot01,kau09,lee06}. Although the more complex 
related so-called misfit cobaltates appear to display even larger thermopower and 
increased figure of merit (see e.g.~\cite{heb13} for a recent review), the sodium 
cobaltate system still holds most of the main physics ready in its simplest structural
form. There have been various theoretical modelings of the Seebeck coefficient for 
this system~\cite{kos01,ham07,kur07,pet07,wis10,san12}, ranging from the use of
Heikes formula, Boltzmann-equation approaches as well as Kubo-formula oriented
modelings. Albeit for a full account of thermoelectricity details may 
matter~\cite{wis10,san12}, for the doping regime 0.6$<$$x$$<$0.75 nearly all different
theoretical descriptions yield thermopower values within the ranges of the experimental
data. However open modeling questions remain for the highly increased Seebeck values
in the regime of vary large doping $x$$>$0.8~\cite{lee06,pet07} as well as for lower 
dopings $x$$\lesssim$0.5, where e.g. a nonmonotonic $S(T)$ with decreasing 
temperature is observed~\cite{kau09}.

Here we exhibit results for the thermopower as obtained within our charge 
self-consistent DFT+DMFT-based Kubo-formalism approach which builds up on the 
$t_{2g}$ correlated subspace for Na$_x$CoO$_2$. Data is provided for $x$=0.7 and
$x$=0.75 in Fig.~\ref{seebeck-pic}, the other more challenging doping regimes 
concerning the thermoelectric response will be addressed in future studies. For 
instance, we expect that the low-lying $e_g'$ bands leave some fingerprints in the 
thermopower for small $x$. Longer-ranged FM spin fluctuations and charge-ordering
tendencies~\cite{pei11} may influence $S(T)$ for $x$$>$0.8. Our inplane Seebeck 
values are in good agreement with experimental data of Kaurav 
{\sl et al.}~\cite{kau09}. The increased response for $x$=0.75 compared to $x$=0.7 is 
also verified, albeit the experimental tendency towards stronger enhancement with 
increased doping is still somewhat underestimated from theory. In addition to the 
inplane values, Fig.~\ref{seebeck-pic} depicts the $S(T)$ tensor part
along the $c$-axis of the system. Besides the $n$-like response with a change of
sign, the absolute value becomes reduced at larger $T$, related to the different 
(rather incoherent) transport between layers at elevated temperature~\cite{val02}.

\section{Two-particle function: dynamic spin susceptibility}
Besides the one-particle spectral properties and the thermopower, a further 
intriguing cobaltate
issue is the magnetic behavior with doping. Within the frustrated triangular 
CoO$_2$ layers superexchange may dominate the low-doping regime due to nominal Mott 
proximity, but competing exchange processes set in at larger doping. The work by 
Lang {\sl et al.}~\cite{lan08} based on nuclear-magnetic-resonance measurements nicely 
summarized the magnetic phase diagram of Na$_x$CoO$_2$ with temperature $T$, showing 
the inplane crossover from antiferromagnetic (AFM) to ferromagnetic (FM) correlations 
with the eventual onset of A-type AFM order for $x$$>$0.75.

In line with the results for the spectral function we computed the spin susceptibility
$\chi_s(\omega,{\bf q},T)$ for an effective single-band model using DMFT with local 
vertex contributions~\cite{boe12}. This allows for the theoretical verification of
the AFM-to-FM crossover. It can directly be retrieved from the shift of maxima in 
the static part 
$\chi_s(\omega$=$0,{\bf q},T)$ for ${\bf q}$ at the BZ $K$ point at small $x$ towards 
maxima at ${\bf q}$=0 ($\Gamma$ point) at larger doping.
Figure~\ref{dynsus} displays the full dynamic spin susceptibility with increasing $x$
in the paramagnetic regime. Below $x$=0.5 the strong two-particle spectral 
intensity close to $M$ and $K$ at the BZ boundary is indeed visible. For rather large 
doping the intensity accumulates at small frequency $\omega$ near the $\Gamma$ point, 
with clear paramagnon branches due to the proximity towards inplane FM order. We note
that the vertex contributions are essential for the qualitative as well as quantitative
signatures in the doping-dependent spin susceptibility~\cite{boe12}.

Most interestingly, there also is a high-intensity mode near the $K$ point 
with maximum spectral weight well located around the comensurable doping $x$=0.67
on the frustrated CoO$_2$ triangular lattice. The corresponding 
excitation energy of about 1 eV for $T$=580K is decreasing with lowering the
temperature. Thus albeit the low-energy spin excitations for that larger doping
have already shifted towards FM kind, a rather stable finite-$\omega$ AFM-like mode 
becomes available. This intriguing doping and frequency dependence of the effective 
exchange $J$ can be linked to the specific hopping-integral structure of sodium
cobaltate. In this respect Fig.~\ref{dynsus_tb} shows the dynamic spin susceptibility
at $x$=0.67 for the Hubbard model on the triangular lattice with only nearest-neighbor
hopping $-t$. While the FM paramagnon modes close to $\Gamma$ seem even 
strengthened in that case, the high-energy feature close to $K$ is now 
completely absent. Still it is not obvious to draw a straightforward connection 
between the one-particle spectral function and two-particle dynamic spin 
susceptibility.

\section{Summary}
We have presented a state-of-the-art DFT+DMFT investigation of the multi-orbital
one-particle spectral properties as well as the 
thermoelectric behavior of Na$_x$CoO$_2$.
The charge self-consistent scheme brings the one-particle spectral function 
concerning the correlated Fermi surface and the broadening of the occupied part in
good agreement with available ARPES data. 
Further extensions of the realistic methodology towards the proper 
inclusion of charge-order effects, eventually with incorporating relevant intersite
Coulomb terms, are still needed for a comprehensive understanding. Nonetheless the
present framework is capable of addressing the temperature- and doping-dependent
thermopower in line with experimental data for the larger doping regime. Detailed
studies of the more critical regions in this respect at low and very high dopings
are envisaged. Through the inclusion of sophisticated vertex contributions in a
simplified tight-binding-based approach, details
of the dynamic spin susceptibility, e.g. the prediction of a rather stable
high-energy AFM-like mode close to $x$=0.67, have been revealed. 
Additional experimental work is needed to 
verify our results and to stimulate future work. Eventually, the investigation
of the direct impact of two-particle correlations on the one-particle spectrum is
a challenging goal, but therefore it will be necessary to go beyond the 
local-correlation viewpoint of DMFT.

\begin{acknowledgement}
We thank D. Grieger, A. I. Lichtenstein and O. E. Peil for helpful discussions. 
Financial support from the DFG-SPP1386 and the DFG-FOR1346 is acknowledged. 
Computations were performed at the local computing center of the University of 
Hamburg as well as the North-German Supercomputing Alliance (HLRN) under the grant 
hhp00026.
\end{acknowledgement}

\bibliographystyle{pss}
\bibliography{bibextra}

\end{document}